\documentclass[amstex,thmsa, 10pt]{article}
\usepackage[tbtags]{amsmath}
\usepackage[]{amssymb}
\usepackage[]{amsbsy}
\usepackage[]{amsthm}
\usepackage{graphicx}
\usepackage{pgfplots}
\usepackage{tikz}

\newtheorem{theorem}{Theorem} %[section]

\newtheorem{example}{Example}% [section]

\def\be{\begin{equation}}
\def\ee{\end{equation}}
\def\bea{\begin{eqnarray}}
\def\eea{\end{eqnarray}}

\begin{document}

%\baselineskip 18pt
%\linenumbers

\title{\Large \textbf{Simple Proofs of the Summation and Connectivity Theorems in Metabolic Control
Analysis }}

\author{Weijiu Liu\thanks{Corresponding author. Email:
weijiul@uca.edu, Phone: 1-501-450-5661,
 Fax: 1-501-450-5662}      \\
\normalsize{ Department of  Mathematics }\\
 \normalsize{  University of Central Arkansas }\\
  \normalsize{201 Donaghey Avenue, Conway, AR 72035, USA }\\
   }

\date{}
 \maketitle

 \begin{abstract}
 In the early 1970s,    the Kacser/Burns   and the Heinrich/Rapoport groups
 independently discovered    the   important summation
 and connectivity theorems
 in metabolic control analysis.    These theorems  were derived originally
    by using
     thought experiments and proved mathematically later.
     The mathematical proofs are not easy for me
      to read and follow. But
     the proofs actually can be very simple and need only a couple
     of lines  as I give here.
 \end{abstract}

 \textbf{Mathematics Subject Classification (2020)}: 92C42, 34A34

\section{Introduction}
In the early 1970s, two groups,  the Kacser/Burns group \cite{Kacser-1973} and the Heinrich/Rapoport group \cite{Heinrich-1974},
 independently discovered    the   important summation
 and connectivity theorems
 in metabolic control analysis. These theorems were derived  in the original
   papers \cite{Kacser-1973} and \cite{Heinrich-1974}
    by using
     thought experiments
    \cite[page 170]{Sauro-book-2013}  on metabolic networks, imagining a scenario to explore concepts and propose equations and relations from which the theorems  can be derived.
 Later they were proved mathematically (see, e.g.,
 \cite[Chapter 5]{Heinrich-book-1996}).
 The mathematical proofs are not easy for me
  to read and follow. But actually
     the   proofs   can be very simple and need only a couple
     of lines as I give here.

\section{The Proof  of the Summation Theorem}
The summation theorem can be stated in a    general
mathematical form
 as follows.

  \begin{theorem} \label{sum-1-theorem}
  Let
  \begin{equation}
       \mathbf{F}(\mathbf{e}, \mathbf{S})=
    \mathbf{0},
   \label{F-func}
 \end{equation}
  and
  \begin{equation*}%\label{J-func}
  J=J(\mathbf{e}, \mathbf{S}),
   \end{equation*}
  where  $\mathbf{e} = (e_1, e_2, \cdots, e_n)$,
  $\mathbf{S} = (S_1, S_2, \cdots, S_m)$,    $J$ is a differentiable function from $\mathbb{R}^n\times \mathbb{R}^m\to\mathbb{R}$, and
  $$
  \mathbf{F}(\mathbf{e}, \mathbf{S}) =\left(\begin{array}{c}
                            F_1(\mathbf{e}, \mathbf{S}) \\
                           \vdots  \\
                            F_m(\mathbf{e}, \mathbf{S})
                          \end{array}\right)  $$
                          is   a differentiable vector function from $\mathbb{R}^n\times \mathbb{R}^m\to\mathbb{R}^m$.
   Assume that the Jacobian matrix
  \begin{eqnarray*}
    \frac{\partial \mathbf{F} }{\partial \mathbf{S}} &=& \left(\begin{array}{ccc}
             \frac{\partial F_1}{\partial S_1} &\cdots & \frac{\partial F_1}{\partial S_m}\\
             \vdots&\cdots & \vdots\\
             \frac{\partial F_m}{\partial S_1} &\cdots & \frac{\partial F_m}{\partial S_m}
           \end{array}
           \right)  %\label{Jacbian-m-dim}
  \end{eqnarray*}
  is nonsingular.
  Then
  \begin{eqnarray}
   \frac{\partial \mathbf{S}}{\partial \mathbf{e}}
        &=&-\left( \frac{\partial \mathbf{F}}{\partial \mathbf{S}}\right)^{-1}  \frac{\partial \mathbf{F}}{\partial \mathbf{e}}, \label{general-sum-0}\\
        \frac{\partial J}{\partial \mathbf{e}}
     &=&  -\frac{\partial J}{\partial \mathbf{S}}
     \left( \frac{\partial \mathbf{F}}{\partial \mathbf{S}}\right)^{-1}  \frac{\partial \mathbf{F}}{\partial \mathbf{e}}+ \frac{\partial J}{\partial \mathbf{e}}.
     \label{general-sum-1}
  \end{eqnarray}
  Note that $\mathbf{S}$ is treated as an independent variable in $ \frac{\partial J}{\partial \mathbf{e}}$ on
  the right hand side of the above equation.
  Furthermore, if
  \begin{eqnarray}
  \sum_{j=1}^{n} e_j\frac{\partial F_i}{\partial e_j} &=& 0,
  \quad i = 1, \cdots, m,
 \label{sum-0-condition}\\
 \sum_{j=1}^{n} e_j\frac{\partial J}{\partial e_j} &=& J,
 \label{sum-1-condition}
  \end{eqnarray}
 then
  \begin{equation}\label{sum-to-0}
    \sum_{j=1}^{n}e_j\frac{\partial S_i}{\partial e_j} =0,\quad
  i = 1,\cdots, m
  \end{equation}
  and
  \begin{equation}\label{sum-to-1}
   \sum_{j=1}^{n}e_j\frac{\partial J}{\partial e_j} =J.
  \end{equation}
\end{theorem}

  \textit{Proof}.  Since the Jacobian matrix $ \frac{\partial \mathbf{F}}{\partial \mathbf{S}}$
    is assumed to be nonsingular,
    the implicit
function theorem (see, e.g., \cite[page 148]{Deimling-book-1985}) ensures that the system \eqref{F-func}
can be solved for $S_1, \cdots, S_m$ to get   $S_i = S_i(e_1, e_2, \cdots, e_n)$, $i = 1, 2, \cdots,
m$.
  Using the chain rule to differentiate the equations \eqref{F-func}  in $\mathbf{e}$, we obtain that
   \begin{equation*}%\label{diff-F-system}
      \frac{\partial \mathbf{F}}{\partial \mathbf{S}}
     \frac{\partial \mathbf{S}}{\partial \mathbf{e}}+ \frac{\partial \mathbf{F}}{\partial \mathbf{e}}
        =\mathbf{0}.
  \end{equation*}
  Since the Jacobian matrix $ \frac{\partial \mathbf{F} }{\partial \mathbf{S} }  $
    is assumed to be nonsingular, we derive that
    \begin{equation*}
     \frac{\partial \mathbf{S}}{\partial \mathbf{e}}
        =-\left( \frac{\partial \mathbf{F}}{\partial \mathbf{S}}\right)^{-1}  \frac{\partial \mathbf{F}}{\partial \mathbf{e}}.
  \end{equation*}
  This proves \eqref{general-sum-0}.

  Using the chain rule to differentiate $J $ in $\mathbf{e}$ and
  using the equation \eqref{general-sum-0}, we obtain that
  \begin{equation*}%\label{diff-J }
   \frac{\partial J}{\partial \mathbf{e}}=  \frac{\partial J}{\partial \mathbf{S}}
     \frac{\partial \mathbf{S}}{\partial \mathbf{e}}+ \frac{\partial J}{\partial \mathbf{e}}
     =  -\frac{\partial J}{\partial \mathbf{S}}
     \left( \frac{\partial \mathbf{F}}{\partial \mathbf{S}}\right)^{-1}  \frac{\partial \mathbf{F}}{\partial \mathbf{e}}+ \frac{\partial J}{\partial \mathbf{e}}.
  \end{equation*}
  Note that $\mathbf{S}$ is treated as an independent variable in $ \frac{\partial J}{\partial \mathbf{e}}$ on
  the right hand side of the above equation.
  This proves the equation \eqref{general-sum-1}.

  If the conditions \eqref{sum-0-condition} and \eqref{sum-1-condition} are satisfied, then, multiplying the
  equations \eqref{general-sum-0} and \eqref{general-sum-1}
  by $\mathbf{e}$, one can readily derive the equations \eqref{sum-to-0} and \eqref{sum-to-1}.

 \section{The Proof of the Connectivity Theorem}.

 The connectivity theorem can be stated in a    general
mathematical form
 as follows.
\begin{theorem} \label{connectivity-theorem} Let
\begin{equation}
       \mathbf{F}(\mathbf{v}(\mathbf{e}, \mathbf{S}))=
    \mathbf{0},
   \label{equi-system}
 \end{equation}
  where  $\mathbf{e} = (e_1, e_2, \cdots, e_n)$,
  $\mathbf{S} = (S_1, S_2, \cdots, S_m)$,
  $$\mathbf{v} (\mathbf{e}, \mathbf{S})=\left(\begin{array}{c}
                            v_1(\mathbf{e}, \mathbf{S}) \\
                           \vdots  \\
                            v_n(\mathbf{e}, \mathbf{S})
                          \end{array}\right) $$
 is a differentiable vector function from $\mathbb{R}^n\times \mathbb{R}^m\to\mathbb{R}^n$,
     and
  $$
  \mathbf{F}(\mathbf{v}(\mathbf{e}, \mathbf{S})) =\left(\begin{array}{c}
                            F_1(\mathbf{v}(\mathbf{e}, \mathbf{S})) \\
                           \vdots  \\
                            F_m(\mathbf{v}(\mathbf{e}, \mathbf{S}))
                          \end{array}\right)  $$
                          is   a differentiable vector function from $\mathbb{R}^n \to\mathbb{R}^m$.
 Assume that the Jacobian matrices
  \begin{eqnarray*}
     \frac{\partial   \mathbf{F} }{\partial \mathbf{S} }   &=& \frac{\partial   \mathbf{F} }{\partial \mathbf{v} }\frac{\partial   \mathbf{v} }{\partial \mathbf{S} }=
     \left(\begin{array}{ccc}
             \frac{\partial F_1}{\partial v_1} &\cdots & \frac{\partial F_1}{\partial v_n}\\
             \vdots&\cdots & \vdots\\
             \frac{\partial F_m}{\partial v_1} &\cdots & \frac{\partial F_m}{\partial v_n}
           \end{array}
           \right)
      \left(\begin{array}{ccc}
             \frac{\partial v_1}{\partial S_1} &\cdots & \frac{\partial v_1}{\partial S_m}\\
             \vdots&\cdots & \vdots\\
             \frac{\partial v_n}{\partial S_1} &\cdots & \frac{\partial v_n}{\partial S_m}
           \end{array}
           \right) ,\\ %\label{nonsingular-Jacbian-A}\\
           \frac{\partial \mathbf{v}}{\partial \mathbf{e}}
            &=&  \left(\begin{array}{ccc}
             \frac{\partial v_1}{\partial e_1} &\cdots & \frac{\partial v_1}{\partial e_n}\\
             \vdots&\cdots & \vdots\\
             \frac{\partial v_n}{\partial e_1} &\cdots & \frac{\partial v_n}{\partial e_n}
           \end{array}
           \right) ,%\label{nonsingular-Jacbian-v}
  \end{eqnarray*}
  are nonsingular.
  Then
  \begin{eqnarray}
    \frac{\partial \mathbf{S}}{\partial \mathbf{e}} \left( \frac{\partial \mathbf{v}}{\partial \mathbf{e}}\right)^{-1}
     \frac{\partial \mathbf{v}}{\partial \mathbf{S}}&=& -\mathbf{I},\label{S-connectivity} \\
      \frac{\partial \mathbf{v}}{\partial \mathbf{e}} \left( \frac{\partial \mathbf{v}}{\partial \mathbf{e}}\right)^{-1}
     \frac{\partial \mathbf{v}}{\partial \mathbf{S}}&=&  \mathbf{0},\label{J-connectivity}
  \end{eqnarray}
  where $\mathbf{I}$ is the identity matrix.
  Note that $\mathbf{S}$ is treated as an independent variable in $ \left(\frac{\partial \mathbf{v}}{\partial \mathbf{e}}\right)^{-1}$
  in the above equation.
  Furthermore, if
 \begin{eqnarray}
   \frac{\partial v_i}{\partial e_j} &=& 0\quad \mbox{if }\; i\neq j, \label{connectivity-cond1} \\
    e_i \frac{\partial v_i}{\partial e_i} &=&v_i, \quad  i=1,\cdots, n,\label{connectivity-cond2}
 \end{eqnarray}
 then
  \begin{eqnarray}
     \sum_{k=1}^{n} \frac{e_k}{v_k}
      \frac{\partial S_i}{\partial e_k}
      \frac{\partial v_k}{\partial S_j}&=&\left\{
      \begin{array}{ll}
        -1 & \mbox{if }\; i=j, \\
        0 & \mbox{if }\; i\neq j,
      \end{array} \right.\label{S-connectivity-special} \\
      \sum_{k=1}^{n} \frac{e_k}{v_k}
      \frac{\partial v_i}{\partial e_k}
      \frac{\partial v_k}{\partial S_j}&=&0.\label{J-connectivity-special}
  \end{eqnarray}

\end{theorem}

\textit{Proof}.  Since the Jacobian matrix $ \frac{\partial \mathbf{F} }{\partial \mathbf{S} } $
    is assumed to be nonsingular,
    the implicit
function theorem (see, e.g., \cite[page 148]{Deimling-book-1985}) ensures that the system \eqref{equi-system}
can be solved for $S_1, \cdots, S_m$ to get   $S_i = S_i(e_1, e_2, \cdots, e_n)$, $i = 1, 2, \cdots,
m$.
  Using the chain rule to differentiate the system \eqref{equi-system} in $\mathbf{e}$, we obtain that
\begin{equation}\label{diff-equi-system}
    \frac{\partial   \mathbf{F} }{\partial \mathbf{v} }\frac{\partial \mathbf{v}}{\partial \mathbf{S}}
     \frac{\partial \mathbf{S}}{\partial \mathbf{e}}+\frac{\partial   \mathbf{F} }{\partial \mathbf{v} }\frac{\partial \mathbf{v}}{\partial \mathbf{e}}
        =\mathbf{0}.
  \end{equation}
Since the Jacobian matrix $ \frac{\partial \mathbf{F} }{\partial \mathbf{S} } $
    is assumed to be nonsingular, we derive that
    \begin{equation*}
     \frac{\partial \mathbf{S}}{\partial \mathbf{e}}
        =-\left(\frac{\partial   \mathbf{F} }{\partial \mathbf{v} }\frac{\partial \mathbf{v}}{\partial \mathbf{S}}\right)^{-1} \frac{\partial   \mathbf{F} }{\partial \mathbf{v} }\frac{\partial \mathbf{v}}{\partial \mathbf{e}}.
  \end{equation*}
  Multiplying this equation by $ \left(\frac{\partial \mathbf{v}}{\partial \mathbf{e}}\right)^{-1}\frac{\partial \mathbf{v}}{\partial \mathbf{S}}$
  gives that
  \begin{equation*}
     \frac{\partial \mathbf{S}}{\partial \mathbf{e}}
     \left(\frac{\partial \mathbf{v}}{\partial \mathbf{e}}\right)^{-1}\frac{\partial \mathbf{v}}{\partial \mathbf{S}}
        =-\left(\frac{\partial   \mathbf{F} }{\partial \mathbf{v} }\frac{\partial \mathbf{v}}{\partial \mathbf{S}}\right)^{-1} \frac{\partial   \mathbf{F} }{\partial \mathbf{v} }  \frac{\partial \mathbf{v}}{\partial \mathbf{e}} \left(\frac{\partial \mathbf{v}}{\partial \mathbf{e}}\right)^{-1}\frac{\partial \mathbf{v}}{\partial \mathbf{S}} = -\mathbf{I}.
  \end{equation*}
  This proves \eqref{S-connectivity}.

  Using the chain rule to differentiate $\mathbf{v} $ in $\mathbf{e}$, we obtain that
  \begin{equation*}%\label{diff-v }
   \frac{\partial \mathbf{v}}{\partial \mathbf{e}}=  \frac{\partial \mathbf{v}}{\partial \mathbf{S}}
     \frac{\partial \mathbf{S}}{\partial \mathbf{e}}+ \frac{\partial \mathbf{v}}{\partial \mathbf{e}}.
  \end{equation*}
  Note that $\mathbf{S}$ is treated as an independent variable in $ \frac{\partial \mathbf{v}}{\partial \mathbf{e}}$ on
  the right hand side of the above equation.
  Multiplying this equation by $
  \left(\frac{\partial \mathbf{v}}{\partial \mathbf{e}}\right)^{-1}\frac{\partial \mathbf{v}}{\partial \mathbf{S}}$ and using
  the equation \eqref{J-connectivity},
  we obtain that
\begin{equation*}
   \frac{\partial \mathbf{v}}{\partial \mathbf{e}}
   \left(\frac{\partial \mathbf{v}}{\partial \mathbf{e}}\right)^{-1}\frac{\partial \mathbf{v}}{\partial \mathbf{S}}
   =  \frac{\partial \mathbf{v}}{\partial \mathbf{S}}
     \frac{\partial \mathbf{S}}{\partial \mathbf{e}}
     \left(\frac{\partial \mathbf{v}}{\partial \mathbf{e}}\right)^{-1}\frac{\partial \mathbf{v}}{\partial \mathbf{S}} + \frac{\partial \mathbf{v}}{\partial \mathbf{e}}
     \left(\frac{\partial \mathbf{v}}{\partial \mathbf{e}}\right)^{-1}\frac{\partial \mathbf{v}}{\partial \mathbf{S}}
     =  -\frac{\partial \mathbf{v}}{\partial \mathbf{S}}
      +  \frac{\partial \mathbf{v}}{\partial \mathbf{S}}  =\mathbf{0}.
  \end{equation*}
  Note that $\mathbf{S}$ is treated as an independent variable in $ \left(\frac{\partial \mathbf{v}}{\partial \mathbf{e}}\right)^{-1}$
  in the above equation.
  This proves the equation \eqref{J-connectivity}.

 If the conditions \eqref{connectivity-cond1} and \eqref{connectivity-cond2} are satisfied, then $
 \left(\frac{\partial \mathbf{v}}{\partial \mathbf{e}}\right)^{-1}$
 becomes a simple diagonal matrix:
 $$\left(\frac{\partial \mathbf{v}}{\partial \mathbf{e}}\right)^{-1} =
 \left(
 \begin{array}{cccc}
   \left(\frac{\partial v_1}{\partial e_1}\right)^{-1} &
   0 & \cdots&
   0 \\
  0 &
   \left(\frac{\partial v_2}{\partial e_2}\right)^{-1} & \cdots&
  0 \\
   \vdots &\vdots & \cdots&\vdots\\
  0&
   0 & \cdots&
   \left(\frac{\partial v_n}{\partial e_n}\right)^{-1}
 \end{array}
 \right)
 =
 \left(
 \begin{array}{llll}
    \frac{e_1}{v_1}  &
   0 & \cdots&
    0 \\
     0  &
   \frac{e_2}{v_2}  & \cdots&
    0 \\
   \vdots &\vdots & \cdots&\vdots\\
   0  &
   0 & \cdots&
    \frac{e_n}{v_n}
 \end{array}
 \right),$$
 and therefore, the equations \eqref{S-connectivity} and
 \eqref{J-connectivity} become the equations
 \eqref{S-connectivity-special} and \eqref{J-connectivity-special},
 respectively.

\section{Examples}

 When applying the above theorems   to a   metabolic pathway,
  $e_j$ stand for the concentrations of enzymes,
$S_i$ for the concentrations of substrates,   $e_j\frac{\partial S_i}{\partial e_j}/S_i$ for the concentration control coefficients, and $e_j\frac{\partial J}{\partial e_j}/J $ for the flux control coefficients. The system \eqref{F-func} or \eqref{equi-system} is the equilibrium equations
for a metabolic network.

\begin{example}
     Consider a simple branched  metabolic pathway as shown
 in Figure \ref{pathway}. Assume that $S_1$, $S_4$, $S_5$, and $S_6$ are held fixed.
\begin{figure}[h]
  \begin{center}
\begin{tikzpicture}[scale = 1]
\draw[line width=0.5mm,  ->]  (0,0)  --  (2,0);
\draw[line width=0.5mm, <- ]  (0,0)  --  (2,0);
  \draw (1,0)node[above] {$ \mathbf{v_1}$}  ;
  \draw (1,0)node[below] {$ \mathbf{E_1}$}  ;
 \draw (0,0)node[left] {$ \mathbf{S_1}$}  ;
 \draw (2,0)node[right] {$ \mathbf{S_2}$}  ;

 \draw[line width=0.5mm,  ->]  (2.3,-0.3)  --  (2.3,-2.3);
\draw[line width=0.5mm, <- ]   (2.3,-0.3)  --  (2.3,-2.3);
  \draw (2.3,-1.3)node[right] {$ \mathbf{v_4}$}  ;
  \draw (2.3,-1.3)node[left] {$ \mathbf{E_4}$}  ;
 \draw (2.3,-2.3)node[below] {$ \mathbf{S_5}$}  ;

 \draw[line width=0.5mm,  ->]  (2.7,0)  --  (4.7,0);
 \draw[line width=0.5mm,  <- ]  (2.7,0)  --  (4.7,0);
 \draw (3.7,0)node[above] {$ \mathbf{v_2}$}  ;
  \draw (3.7,0)node[below] {$ \mathbf{E_2}$}  ;
 \draw (4.7,0)node[right] {$ \mathbf{S_3}$}  ;

 \draw[line width=0.5mm,  ->]  (5, 0.3)  --  (5, 2.3);
\draw[line width=0.5mm, <- ]   (5, 0.3)  --  (5, 2.3);
  \draw (5, 1.3)node[left] {$ \mathbf{v_5}$}  ;
  \draw (5, 1.3)node[right] {$ \mathbf{E_5}$}  ;
 \draw (5, 2.3)node[above] {$ \mathbf{S_6}$}  ;

 \draw[line width=0.5mm,  ->]  (5.4,0)  --  (7.4,0);
 \draw[line width=0.5mm,  <- ]  (5.4,0)  --  (7.4,0);
 \draw (6.4,0)node[above] {$ \mathbf{v_3}$}  ;
  \draw (6.4,0)node[below] {$ \mathbf{E_3}$}  ;
 \draw (7.4,0)node[right] {$ \mathbf{S_4}$}  ;

 \end{tikzpicture}

  \end{center}
 \caption{A simple   metabolic pathway. $S_i$
 stand for substrates, $E_i$   for enzymes, and
 $v_i$  for the reaction rates.}
 \label{pathway}
\end{figure}
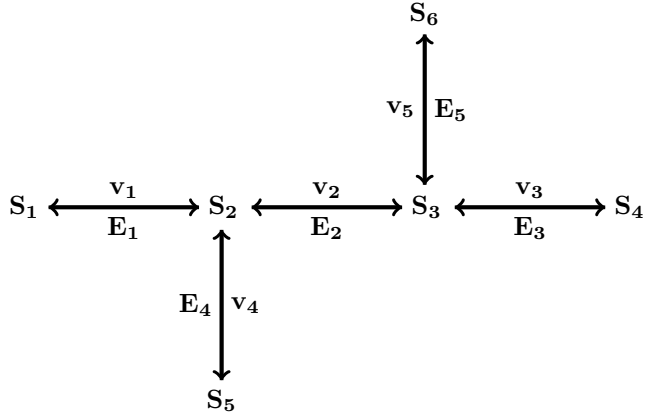

\end{example}

  The reaction equations are given as follows:
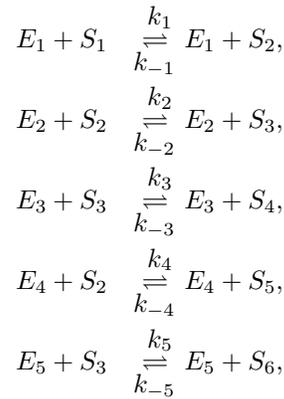
\begin{eqnarray*}
E_1+S_1 \hspace{0.3em}&
   \rightleftharpoons & \hspace{- 1.8em}
\raisebox{2.0ex}{$k_1$} \hspace{-1.5em}\raisebox{-2.0ex}{$k_{-1}$ }
E_1+S_2,\\
   E_2+S_2 \hspace{0.3em}&  \rightleftharpoons& \hspace{-1.8em}
\raisebox{2.0ex}{$k_2$} \hspace{-1.5em}\raisebox{-2.0ex}{$k_{-2}$ }E_2+S_3,\\
  E_3+S_3 \hspace{0.3em}&  \rightleftharpoons& \hspace{-1.8em}
\raisebox{2.0ex}{$k_3$} \hspace{-1.5em}\raisebox{-2.0ex}{$k_{-3}$
}E_3+S_4,\\
E_4+S_2 \hspace{0.3em}&  \rightleftharpoons& \hspace{-1.8em}
\raisebox{2.0ex}{$k_4$} \hspace{-1.5em}\raisebox{-2.0ex}{$k_{-4}$
}E_4+S_5,\\
E_5+S_3 \hspace{0.3em}&  \rightleftharpoons& \hspace{-1.8em}
\raisebox{2.0ex}{$k_5$} \hspace{-1.5em}\raisebox{-2.0ex}{$k_{-5}$
}E_5+S_6,
 \end{eqnarray*}
where $k_i$ and $k_{-i}$ are forward and backward
reaction rate constants.
By the mass-action law, the reaction rate of
 each enzyme $E_i$ is
  \begin{eqnarray*}
   J_1 &=&v_1= k_1e_1S_1-k_{-1}e_1S_2,\\
   J_2 &=&v_2 =k_2e_2S_2-k_{-2}e_2S_3,\\
   J_3 &=&v_3= k_3e_3S_3-k_{-3}e_3S_4 ,\\
   J_4 &=&v_4 =k_4e_4S_2-k_{-4}e_4S_5 ,\\
   J_5 &=&v_5= k_5e_5S_3-k_{-5}e_5S_6 .
  \end{eqnarray*}
Then
 the dynamic
 mathematical model
for the metabolic pathway is given by
\begin{eqnarray*}
\frac{dS_2}{dt}&= & k_1e_1S_1-k_{-1}e_1S_2-k_2e_2S_2+k_{-2}e_2S_3
-k_4e_4S_2+k_{-4}e_4S_5 =v_1-v_2-v_4,\\
\frac{dS_3}{dt}&= & k_2e_2S_2-k_{-2}e_2S_3-k_3e_3S_3+k_{-3}e_3S_4
-k_5e_5S_3+k_{-5}e_5S_6=v_2-v_3-v_5,
 \end{eqnarray*}
 or in the matrix form
 \begin{equation*}
   \frac{d}{dt}\left(\begin{array}{l}
                       S_2 \\
                       S_3
                     \end{array}\right)
                     =\left(\begin{array}{rrrrr}
                       1&-1&0&-1&0 \\
                       0&1&-1&0&-1
                     \end{array}\right)
                     \left(\begin{array}{l}
                       v_1  \\
                       v_2  \\
                       v_3  \\
                       v_4  \\
                       v_5
                     \end{array}\right) =\mathbf{N}\mathbf{v}
 \end{equation*}
 where $S_i$ and $e_i$ stand for the concentrations of $S_i$ and $E_i$,
 respectively,  $ \mathbf{v}  = (v_1, v_2,v_3,v_4,v_5)^T$, and
 $$
 \mathbf{N}=\left(\begin{array}{rrrrr}
                       1&-1&0&-1&0 \\
                       0&1&-1&0&-1
                     \end{array}\right)$$
 is the stoichiometric matrix.
 It is clear that     the equilibrium system
\begin{eqnarray*}
0&= & k_1e_1S_1-k_{-1}e_1S_2-k_2e_2S_2+k_{-2}e_2S_3
-k_4e_4S_2+k_{-4}e_4S_5,\\
0&= & k_2e_2S_2-k_{-2}e_2S_3-k_3e_3S_3+k_{-3}e_3S_4
-k_5e_5S_3+k_{-5}e_5S_6
 \end{eqnarray*}
 or
 $$\mathbf{N}\mathbf{v} = \mathbf{0}$$
 is of the form of the equation \eqref{F-func} or
 \eqref{equi-system} and conditions \eqref{sum-0-condition},
 \eqref{sum-1-condition}, \eqref{connectivity-cond1},
 and \eqref{connectivity-cond2} are satisfied.
 Thus, according to the equation  \eqref{sum-to-0},
the concentration control coefficients sum  to 0:
 $$C^{S_i}_{e_1}+ C^{S_i}_{e_2} +C^{S_i}_{e_3}+C^{S_i}_{e_4}+C^{S_i}_{e_5}=0,\quad i = 2, 3, $$
  where the concentration control coefficient is defined by
  $$C^{S_j}_{e_i}=\frac{e_i}{S_j}\frac{\partial S_j}{\partial e_i}.$$
 According to the equation    \eqref{sum-to-1},  the flux control coefficients
sum  to 1:
$$
C^{J_i}_{e_1}+ C^{J_i}_{e_2} +C^{J_i}_{e_3}+C^{J_i}_{e_4}+C^{J_i}_{e_5}=1,\quad
 i = 1, 2, 3, 4, 5,
$$
where   the flux control coefficient is defined  by
$$C^{J_j}_{e_i} =\frac{e_i}{J_j}\frac{\partial J_j}{\partial e_i}.$$
According to the equation    \eqref{S-connectivity-special},
the concentration connectivity equations hold:
 $$C^{S_i}_{e_1}E^{v_1}_{S_i}+ C^{S_i}_{e_2}E^{v_2}_{S_i} +C^{S_i}_{e_3}E^{v_3}_{S_i}
 +C^{S_i}_{e_4}E^{v_4}_{S_i}+C^{S_i}_{e_5}E^{v_5}_{S_i}=-1,
 \quad i = 2, 3,  $$
where   the elasticity coefficient is defined  by
$$
E^{v_i}_{S_j}=\frac{S_j}{v_i}\frac{\partial v_i}{\partial S_j}
$$
According to the equation    \eqref{J-connectivity-special},
the flux connectivity equations hold:
 $$C^{v_i}_{e_1}E^{v_1}_{S_j}+ C^{v_i}_{e_2}E^{v_2}_{S_j} +C^{v_i}_{e_3}E^{v_3}_{S_j}
 +C^{v_i}_{e_4}E^{v_4}_{S_j}
 +C^{v_i}_{e_5}E^{v_5}_{S_j}=0,\quad
  i=1, 2, 3, 4, 5, j = 2, 3.$$
  Indeed,   these equations can be verified by using the Matlab \cite{MATLAB} or other
  mathematical softwares.

   The pathway in the Figure \ref{pathway} may be treated
 as a traffic flow, $E_i$ being treated as traffic control stations and
 $S_i$ being treated as traffic flows. In this context, Theorem \ref{sum-1-theorem}
 reveals that the change in one traffic control station   will result in the changes in all
 other traffic control stations.

\begin{example}
     Consider an  end-product inhibited metabolic
pathway as shown
 in Figure \ref{inhibited-pathway}.
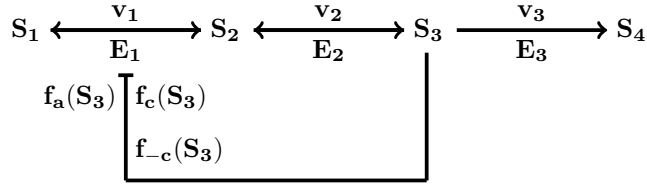
\begin{figure}[h]
  \begin{center}
\begin{tikzpicture}[scale = 1]
\draw[line width=0.5mm,  ->]  (0,0)  --  (2,0);
\draw[line width=0.5mm, <- ]  (0,0)  --  (2,0);
  \draw (1,0)node[above] {$ \mathbf{v_1}$}  ;
  \draw (1,0)node[below] {$ \mathbf{E_1}$}  ;
 \draw (0,0)node[left] {$ \mathbf{S_1}$}  ;
 \draw (2,0)node[right] {$ \mathbf{S_2}$}  ;

 \draw[line width=0.5mm,  ->]  (2.7,0)  --  (4.7,0);
 \draw[line width=0.5mm,  <- ]  (2.7,0)  --  (4.7,0);
 \draw (3.7,0)node[above] {$ \mathbf{v_2}$}  ;
  \draw (3.7,0)node[below] {$ \mathbf{E_2}$}  ;
 \draw (4.7,0)node[right] {$ \mathbf{S_3}$}  ;

 \draw[line width=0.5mm,  ->]  (5.4,0)  --  (7.4,0);
 \draw (6.4,0)node[above] {$ \mathbf{v_3}$}  ;
  \draw (6.4,0)node[below] {$ \mathbf{E_3}$}  ;
\draw (7.4,0)node[right] {$ \mathbf{S_4}$}  ;

 \draw[line width=0.5mm   ]  (5, -0.3)  --  (5, -2);
 \draw[line width=0.5mm   ]  (1, -2)  --  (5, -2);
 \draw[line width=0.5mm   ]  (1, -2)  --  (1, -0.6);
 \draw[line width=0.5mm   ]  (0.9, -0.6)  --  (1.1, -0.6);
  \draw (1, -0.9)node[left] {$ \mathbf{f_a(S_3)}$}  ;
  \draw (1, -0.9)node[right] {$ \mathbf{f_c(S_3)}$}  ;
 \draw (1, -1.6)node[right] {$ \mathbf{f_{-c}(S_3)}$}  ;

 \end{tikzpicture}

  \end{center}
 \caption{An end-product inhibited metabolic
pathway. $S_i$
 stand for substrates, $E_i$   for enzymes, and
 $v_i$  for the reaction rates. The function  $ f_a( S_3)$ models an allosteric
inhibition and the functions $f_c(S_3), f_{-c}(S_3) $ model  a  competitive inhibition.}
 \label{inhibited-pathway}
\end{figure}

\end{example}
Assume that $S_1$ and $S_4$ are held fixed.
The reaction scheme is
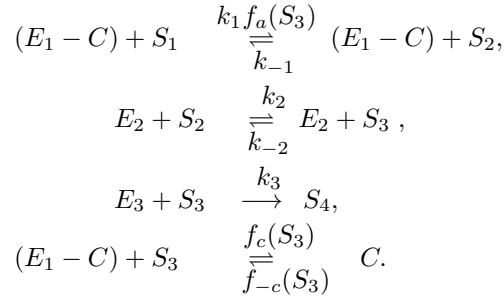
\begin{eqnarray*}
(E_1-C)+S_1 \hspace{1.3em}&
   \rightleftharpoons & \hspace{- 3.5em}
\raisebox{2.0ex}{$k_1f_a(S_3)$} \hspace{-2.4em}\raisebox{-2.0ex}{$k_{-1}$ }
\hspace{1em} (E_1-C)+S_2,\\
  E_2+S_2 \hspace{0.3em}&  \rightleftharpoons& \hspace{-1.8em}
\raisebox{2.0ex}{$k_2$} \hspace{-1.5em}\raisebox{-2.0ex}{$k_{-2}$
}E_2+S_3\hspace{0.3em}    ,\\
 E_3+S_3 \hspace{0.3em}  &\longrightarrow& \hspace{-2em}
\raisebox{2.0ex}{$k_3$} \hspace{0.8em} S_4,\\
(E_1-C)+S_3 \hspace{1.3em}&  \rightleftharpoons& \hspace{-2.5em}
\raisebox{2.0ex}{$f_c(S_3)$} \hspace{-2.8em}\raisebox{-2.0ex}{$f_{-c}(S_3)$
}\hspace{0.8em}C.
 \end{eqnarray*}
Here the function  $ f_a( S_3)$ models an allosteric
inhibition and the functions $f_c(S_3), f_{-c}(S_3), E_1-C$ model  a  competitive inhibition. Using the mass-action law, we
derive the system of the steady states:
\begin{eqnarray}
  k_1f_a( S_3)(e_1-c)S_1-k_{-1}(e_1-c)S_2-k_2e_2S_2+k_{-2}e_2S_3
  &=&0, \label{feedbvack-ode1}\\
  k_2e_2S_2-k_{-2}e_2S_3-k_3e_3S_3-f_c(S_3)(e_1-c)S_3+f_{-c}(S_3)c
  &=&0,\label{feedbvack-ode2}\\
 f_c(S_3)(e_1-c)S_3-f_{-c}(S_3)c&=&0.\label{feedbvack-ode4}
 \end{eqnarray}
 Solving the equation \eqref{feedbvack-ode4} for $c$, we get
 \begin{equation}\label{steady-c}
   c = \frac{  e_1S_3 f_c(S_3) }
   {S_3 f_c(S_3)+f_{-c}(S_3) }.
 \end{equation}
Plugging the solution $c$ into the steady state system,
 we obtain the reduced steady state system
 \begin{eqnarray}
  k_1 e_1S_1f_a( S_3)\left(1-\frac{   S_3 f_c(S_3) }
   {S_3 f_c(S_3)+f_{-c}(S_3) }\right) -k_{-1}e_1S_2\left(1-\frac{  S_3 f_c(S_3) }
   {S_3 f_c(S_3)+f_{-c}(S_3) }\right)
    -k_2e_2S_2+k_{-2}e_2S_3&= &0,
\label{feeback-steady-eq1}\\
  k_2e_2S_2-k_{-2}e_2S_3-k_3e_3S_3 &= &0,\label{feeback-steady-eq2}
 \end{eqnarray}
 The reaction rate of
 each enzyme $E_i$ is
  \begin{eqnarray*}
   v_1 &=&  k_1 e_1S_1f_a( S_3)\left(1-\frac{   S_3 f_c(S_3) }
   {S_3 f_c(S_3)+f_{-c}(S_3) }\right) -k_{-1}e_1S_2\left(1-\frac{  S_3 f_c(S_3) }
   {S_3 f_c(S_3)+f_{-c}(S_3) }\right) ,\\
   v_2   &=&k_2e_2S_2-k_{-2}e_2S_3,\\
   v_3  &=& k_3e_3S_3  .
  \end{eqnarray*}
  Hence, the system \eqref{feeback-steady-eq1} -
  \eqref{feeback-steady-eq2} can be written as
 \begin{eqnarray}
  v_1-v_2 &= &0,\label{stoi-eq1}\\
  v_2-v_3 &= &0.\label{stoi-eq2}
 \end{eqnarray}
 Again,   the system   \eqref{feeback-steady-eq1} -
  \eqref{feeback-steady-eq2} or \eqref{stoi-eq1} - \eqref{stoi-eq2}
  satisfies all conditions of Theorem \ref{sum-1-theorem}
  and Theorem \ref{connectivity-theorem}.
 Therefore, the following summation and connectivity equations
 hold:
 \begin{eqnarray*}
  C^{S_i}_{e_1}+ C^{S_i}_{e_2} +C^{S_i}_{e_3} &=&0,
  \quad i = 2, 3, \\
  C^{v_i}_{e_1}+ C^{v_i}_{e_2} +C^{v_i}_{e_3} &=&1,
  \quad i =1, 2, 3, \\
  C^{S_i}_{e_1}E^{v_1}_{S_i}+ C^{S_i}_{e_2}E^{v_2}_{S_i} +C^{S_i}_{e_3}E^{v_3}_{S_i} &=&-1,
  \quad i = 2, 3, \\
  C^{v_i}_{e_1}E^{v_1}_{S_j}+ C^{v_i}_{e_2}E^{v_2}_{S_j} +C^{v_i}_{e_3}E^{v_3}_{S_j} &=&0,
  \quad i =1, 2, 3, \; j = 2, 3. \\
 \end{eqnarray*}

\end{document}